\documentclass[11pt,a4paper]{article}

\usepackage[T1]{fontenc}
\usepackage[utf8]{inputenc} 
\usepackage{amssymb}
\usepackage{amsmath}
\usepackage{array}
\usepackage{graphicx}
\usepackage[margin = 1in
            ]{geometry}
\usepackage{hhline}
\usepackage{lineno}
\usepackage{booktabs}
\usepackage{subcaption}
\usepackage{xcolor}
\usepackage{hyperref}
\usepackage{natbib}
\usepackage{setspace}
\begin{document}


    \title{Regulatory gray areas of LLM Terms}

    \author{%
    \textbf{Brittany I. Davidson}\\
    University of Bath \\
    Bath, England\\ 
    Email: \href{mailto:bid23@bath.ac.uk}{bid23@bath.ac.uk}
    \and
    \textbf{Kate Muir} \\
    Bath Spa University \\
    Bath, England \\
    Email: \href{mailto:k.muir@bathspa.ac.uk}{k.muir@bathspa.ac.uk}
    \and
    \textbf{Florian A.D. Burnat} \\
    University of Bath \\
    Bath, England \\
    Email: \href{mailto:fadb20@bath.ac.uk}{fadb20@bath.ac.uk}
    \and
    \textbf{Adam N. Joinson} \\
    University of Bath \\
    Bath, England \\ 
    Email: \href{mailto:aj266@bath.ac.uk}{aj266@bath.ac.uk}}

    \date{This version: 
    January 12, 2026 
    \\[.5em]
    Latest version: \href{https://www.florianburnat.com/forward-links/llm_terms/}{Click here}
    }

    \maketitle

    \begin{abstract}
        \noindent Large Language Models (LLMs) are increasingly integrated into academic research pipelines; however, the Terms of Service governing their use remain under-examined. We present a comparative analysis of the Terms of Service of five major LLM providers (Anthropic, DeepSeek, Google, OpenAI, and xAI) collected in November 2025. Our analysis reveals substantial variation in the stringency and specificity of usage restrictions for general users and researchers. We identify specific complexities for researchers in security research, computational social sciences, and psychological studies. We identify `regulatory gray areas' where Terms of Service create uncertainty for legitimate use. We contribute a publicly available resource comparing terms across platforms (OSF) and discuss implications for general users and researchers navigating this evolving landscape. \\
        \vspace{0in}
        
        \bigskip
    \end{abstract}
    
    \setcounter{page}{0}
    \thispagestyle{empty}


\section{Introduction}
Large language models (LLMs) have proliferated in our daily lives and are now integrated across devices and platforms: from email clients that generate and proofread responses, to AI assistants that summarize long texts (e.g., in Adobe software), to academic publisher websites (e.g., ScienceDirect's `Reading Assistant') that summarize articles and answer questions about them. While access to these tools benefits many users, high-profile cases of misuse are increasing. For example, Deloitte has repeatedly used AI in large-scale consultancy work that contained hallucinated references and content \cite{Dhanji2025-vk, Paoli2025-oo}. 

Another harmful example is xAI's Grok, which users employ to generate sexually explicit images of children and women, including feeding women's photos and prompting Grok to undress them without consent \cite{Burgess2026-ew}. Although the tool does not always succeed (sometimes generating images of women in small bikinis instead), this demonstrates that guardrails are failing and users are not following the terms of usage. Alarmingly, little has been done to stop this despite xAI stating they will suspend users for such behavior \cite{Gentleman2026-pt} and the European Commission declaring these images illegal \cite{Sandle2026-xj}. At least 20,000 images have been generated since December 25, 2025, with the number continuing to rise---in theory, another such image is created every few seconds via Grok \cite{Burgess2026-ew}. This highlights both the substantial harm that LLMs can generate and the gaps in how these tools are regulated through company terms of service alongside country and global mandates. 

Here, we focus on the Terms of Service or Usage (henceforth, `Terms') of various LLM tools. Our interest began with recent substantial changes to data and tool access for research as we entered the `post-API' world, where researcher-specific access has been reduced, shifted toward `pay-to-play' models, or simply removed \cite{Freelon2024-yi, Burnat2025-kr}. These API shifts substantially impact research access and integrity; for instance, data sharing for peer review is no longer permitted on some platforms \cite{Davidson2023-uu}. AI and LLMs have sparked new research directions, from technical elements to societal impacts. We observed changes in Terms over time, alongside substantial imbalance in specificity of usage restrictions for general users versus researchers. We analyze the Terms of the most common LLMs and identify issues affecting users and researchers.

\section{Background}
Since their creation and release, interest in the application and impact of LLMs has continued to increase annually \cite{Akhtarshenas2025-cu}. LLMs have demonstrated a variety of abilities across domains via their generation of human-like language, sparking substantial interest across research communities as an `out of the box' advanced AI tool that can engage with a variety of tasks \cite{Xiong2024-sr}. While these new technologies offer opportunities, it is also critically important to understand the limitations and harms that come with them. This has been discussed from a variety of viewpoints, from harms in the training data to noting the issues that will arise should we view LLMs as a tool `with linguistic agency', or to have any agency in themselves (for example, \cite{Birhane2024-po, Bender2021-ct}).

\subsection{Terms \& Conditions}
Terms have a crucial role in informing users about the usage of a tool, product, or service \cite{Luger2013-ru}. They also allow the company to protect itself if something goes `wrong' and a user engages in usage outside of the Terms. Terms are dynamic and naturally require updates. These updates are often shared via notifications/pop-ups, emails, or notes on the policy itself. 


Yet, the Terms for these LLMs are inconsistent and can be vague in some areas (e.g., loose terms around restricted uses). From a researcher's perspective, this raises several issues: researchers may not be aware of changes in Terms that impact their work, or they may need to wrangle research ideas to fit within the Terms if their work falls in a borderline domain. Terms are often both highly specific yet vague (or contradictory), putting researchers in a difficult position of needing to guess, fill in the blanks, and assume that their interpretation aligns with the company's intent. This becomes more complex given that these Terms are updated at random intervals. 

It is also important to note that most companies have notes regarding country-specific Terms, which have other implications for usage and disputes. For example, xAI notes that any disputes will be located in Texas, USA, where the HQ is, whereas DeepSeek notes Hangzhou, China. Others have notes that no matter where a user is located, their Terms override their legal jurisdictions (e.g., Anthropic defines children as under 18 globally). These differences also highlight another challenge of AI training and modeling in a global context.  

\subsection{Who reads the `fine print' anyway?}
It is well known that users tend not to read the Terms in the first place, and it is likely that any updates also remain unread. \citet{Amos2021-fp} found that (privacy) policies change substantially over time by becoming longer and more complex and requiring substantially higher reading levels over the last 20+ years. Hence, it has become a serious question as to how we encourage and help users read Terms and keep updated with them to take an informed opinion as to whether to continue using the service. Furthermore, Terms tend to use complex language that is prohibitively difficult for users to understand and fully consent to \cite{Luger2013-ru}. This is further exacerbated by the way Terms are presented from a user experience perspective, which also has a substantial impact on user awareness, comprehension, and engagement with said Terms \cite{Kitkowska2022-gz, Steinfeld2016-nf}. 

This raises several concerns for individuals and the Terms they're signing up for without fully understanding the implications. This, for example, has been considered more recently with relation to menstrual cycle tracking apps and the inherent threats to privacy bringing harm to women and people who menstruate (e.g., \citet{Dong2022-tn, Cao2024-mj, Zadushlivy2025-sh}). 
For instance, it has been found repeatedly that it is not uncommon for these apps to collect location data, which has no reason to be collected for those tracking their menstrual cycles. This has been highlighted as particularly problematic given legislation changes (e.g., overturning Roe v Wade) in the United States, where the data these apps collect can detect and demonstrate pregnancy and abortion, which is unsafe and puts women at serious risk in the current political climate \cite{Cao2024-mj, Zadushlivy2025-sh}. 

Here, we seek to illuminate the complexities and hidden gray regulatory areas of the Terms of commonly used LLMs. We have two aims. First, we aim to shed light on the implications for researchers across industry and academia that are hidden in the fine print of the Terms. This is because there is a concern that LLM Terms for researchers restrict new knowledge generation regarding the usage and impact of these rapidly accelerating technologies on society. Our second aim is to provide an easily understandable summary of terms to act as a resource and guide for researchers (and users) in this area and provide a resource of cross-referenced Terms (available on the \href{https://osf.io/ypgqk/overview?view_only=9c58559850b84980848a0d5b80c8d00a}{\color{blue}{\underline{OSF}}}).

\section{Methods}
\subsection{Ethical Approval}
We collected the Terms of Service policies from Anthropic, DeepSeek, Google, xAI, and OpenAI. We analyzed these companies in the first instance because they are some of the most popular and well-known LLMs in 2025. We gathered these policies in November 2025 to initiate our analysis. Ethical approval was granted by the Univeristy of Bath [ref = 13448-15424].

\subsection{Approach}
For our analysis, we undertook a document analysis to understand and cross-reference the current Terms across five LLM policies. We chose this method because it allows for rich insights into a context, particularly the contexts in which research participants operate \cite{Bowen2009-nw}. This is particularly important here, as we are using a secondary data analysis of policy documents that have direct downstream implications for all users. Hence, taking a document analysis approach in the first instance allowed us to also integrate content analysis steps by quantifying which companies have Terms relating to specific categories provided (i.e., some Terms had predefined categories within them) and those identified by the authors \cite{Bowen2009-nw}. We felt that this approach that combines methods is particularly important here, as while these data do not `come from' people like an interview or social media posts, but rather, these words gravely impact people, and that is the key for this work is to understand the impacts of these Terms on all users. 

The steps taken were as follows:
\begin{enumerate}
    \item We decided on the five companies of interest, where we placed emphasis on the more prominent and LLMs with large user-bases.
    \item We collected the most up-to-date Terms on their websites. This proved easier for some (e.g., Anthropic) but substantially more complex for others (e.g., Google); we indeed noticed that Google Gemini's Terms were more disparate, and they appeared to only have a general `Generative AI Prohibited Usage Policy' as opposed to a clearly designated Gemini policy. This is perhaps due to the vast number of products and tools Google has and maintains, making it harder to navigate which Terms apply to which tools/platforms/products. 
    \item Based on an initial reading of these Terms, the next step was to begin with the most comprehensive Terms (in this case, Anthropic) and to write out each clause into our `Table of Terms' (available on the \href{https://osf.io/ypgqk/overview?view_only=9c58559850b84980848a0d5b80c8d00a}{\color{blue}{\underline{OSF}}}). Thereafter, each of the other companies’ Terms (Google, xAI, DeepSeek, and OpenAI) was cross-referenced with Anthropic's. Anything Anthroptic's Terms did not have was added into appropriate categories (e.g., illegal activities, child safety, etc)
    \item Once we had obtained all Terms, we analyzed them and categorized them, and adapted them as we generated an overall cross-referenced resource table.
    \begin{enumerate}
        \item This was an abductive process, where we iterated between the Terms of the five companies in order to first see if each set of Terms mentioned an overall category (e.g., `Prohibiting illegal activities' or `Prohibiting users from compromising child safety') and this formed Table \ref{tab:termshighlevel}. These categories were drawn from the categories set in the Terms of Anthropic and OpenAI.
        \item The deeper stage was to iterate between all the Terms across companies to examine which companies had more specific Terms under each category (e.g., within `Prohibiting illegal activities' some companies have a specific clause relating to human trafficking and violation of laws). This formulated the Table of Terms (on the \href{https://osf.io/ypgqk/overview?view_only=9c58559850b84980848a0d5b80c8d00a}{\color{blue}{\underline{OSF}}}), where we show which companies have which Terms. This also allowed us to categorize Terms as we needed to add them. This in particular happened with DeepSeek's Terms, which have a relatively unorganized high-level list of prohibited usages (see Section 3.4) that we added into the respective categories as appropriate. This step also demonstrated at a high level the gaps and vagueness in some companies’ Terms (e.g., with only a few non-specific Terms per category).  
        \item There were sometimes exceptions, that are noted at the bottom of the table (e.g., the minimum age to use the services is often 18, apart from OpenAI that sets a minimum age of 13). 
    \end{enumerate}
    \item The final stage of this analysis was to then analyze these Terms to understand the implications for the general user-base and those in research positions (both in academia and industry). This element, as common with document analysis, drew on content analysis steps to understand how various Terms impact general and research users \cite{Bowen2009-nw}. 
\end{enumerate}

\section{Findings \& Analysis}
\subsection{General Restrictions of Use}
Some providers, particularly xAI, place greater emphasis on ownership of content in their Terms, for example, specifying who owns the content generated, alongside intellectual property rights and data privacy provisions. Most companies have Terms relating to law and regulation, child protection, privacy, disruption of services and critical infrastructure, inciting violence, and harmful content. We have produced a resource available on the \href{https://osf.io/ypgqk/overview?view_only=9c58559850b84980848a0d5b80c8d00a}{\color{blue}{\underline{OSF}}} that compares Terms across the five platforms. Table \ref{tab:termshighlevel} presents a high-level version that provides an overview of the key categories covered in the Terms.

\begin{table}[!h]
\resizebox{\textwidth}{!}{%
\begin{tabular}{llllll}
\toprule
General Categories                                                                  & Anthropic & OpenAI & DeepSeek & Google & xAI \\
 \midrule
Prohibiting illegal activities                                                    & Y         & Y      & Y        & Y      & Y   \\
Prohibiting users to compromise critical infrastructure                           & Y         &        &          &        &     \\
Prohibiting users to compromise computer or network systems                       & Y         & Y      & Y        &        &     \\
Prohibiting users to develop or design weapons                                    & Y         & Y      &          &        &     \\
Prohibiting users from inciting violence or hateful behavior (e.g., terrorist content)                      & Y         & Y      & Y        & Y      & Y   \\
Prohibiting users from compromising privacy or identity                           & Y         & Y      & Y        &        & Y   \\
Prohibiting users from compromising child safety                                  & Y         & Y      & Y        & Y      & Y   \\
Prohibiting users from creating harmful content (e.g., bullying, suicide or pro-eating disorder content)                                   & Y         & Y      & Y        & Y      &     \\
Prohibiting the spread of misinformation                                          & Y         & Y      &          & Y      &     \\
Prohibiting users from undermining democratic processes and political contexts    & Y         & Y      &          & Y      &     \\
Prohibiting criminal justice, censorship, surveillance, law enforcement purposes & Y         & Y      &          &        &     \\
Prohibiting fraud, abuse, predatory practices                                     & Y         & Y      &          & Y      & Y   \\
Prohibiting platform abuse                                                        & Y         & Y      & Y        & Y      & Y   \\
Prohibiting sexually explicit content creation                                    & Y         &        & Y        & Y      &     \\
High-Risk use case requirements                                                   & Y         & Y      &          & Y      & Y   \\
 \bottomrule
\end{tabular}
}
\caption{Overview of Categories within Terms (larger version with every term on \href{https://osf.io/ypgqk/overview?view_only=9c58559850b84980848a0d5b80c8d00a}{\color{blue}{\underline{OSF}}})}
\label{tab:termshighlevel}
\end{table}

From the outset, Anthropic has the most extensive Terms, followed by OpenAI. At this higher level, Google appears to cover many areas; however, within each category, Google has only a few specific clauses within its Generative AI Prohibited Use Policy. Similarly, DeepSeek has fewer Terms that are not overly specific. All companies have some comments on illegal activities, with a vast amount of variance in examples of what this means. Some mention critical infrastructure (Anthropic) and computer network systems (Anthropic, OpenAI, and DeepSeek). These Terms are often related to the use of their services to harm \textit{other} companies’ services, but some also specify their services as prohibited use. Less commonly, there was discussion of the use of services for the creation or development of weapons (Anthropic and OpenAI). It was not uncommon to see other specific `platform abuse' Terms, where all companies have some mention of this. This often included Terms regarding breaching guardrails, circumventing bans, and the co-ordination of malicious activities.

All five state child protection-oriented Terms, which vary substantially in depth, with Anthropic and OpenAI being the most extensive. For example, OpenAI mentions that their `services are designed to prevent harm' under their \textbf{Keep Minors Safe} section, where they list specific restrictions, such as services that cannot be used for grooming, generation of imagery pertaining to children/minors, dangerous challenges for minors, or discussion of shaming related to body type and appearances of minors. This is mirrored by Anthropic's Terms relating to children's safety. 
The prohibitive use policies of xAI, DeepSeek, and Google’s generative AI are much more minimal and broad. DeepSeek additionally notes vague age references: and `\textit{age of 18 or the minimum age required in your country}', which is in contrast to Anthropic, where children or minors are defined as 18 regardless of jurisdiction. Some Terms related to child protection overlap with some sections regarding the generation of harmful content. However, regardless of age, these Terms commonly had specific examples, such as eating disorder (or disordered eating) content, self-harm, or suicide content (Anthropic, OpenAI, and Google). All but xAI note bullying, humiliating, or shaming as prohibited uses. Somewhat adjacently, three companies had specific Terms relating to sexually explicit content (Anthropic, DeepSeek, and Google). This varied in depth, but it was broadly noted that users cannot use their services for sexual role-play or graphic sexual content generation. 

All Terms mention that users are prohibited from inciting violence or hateful behavior (including terrorism), with Anthropic being the most extensive. Most are fairly vague, except for Anthropic, which has much more specific Terms on this issue. There was a separate set of Terms that specifically focused on creating harmful content, which denotes specific content containing shaming, humiliation, bullying, suicide or self-harm content, pro-eating disorder content, animal cruelty, etc.. Anthropic had the most specific coverage, followed by OpenAI, Google, and DeepSeek. xAI did not specifically mention this type of content. 

Only Anthropic, OpenAI, and Google specifically mentioned that their services cannot be used for the spread of misinformation. Anthropic is the most specific, including Terms relating to specific content misrepresenting laws and regulations, generating misleading content with the intention of targeting groups or individuals, or generating false or misleading health, medical, and science-related content. Google was light touch, as was OpenAI; however, Google does specify generating content and claiming it was made by a human is specifically prohibited with their services. Adjacently, Anthropic, OpenAI, and Google all have specific Terms relating to undermining democratic and political content, with Anthropic being the most detailed, covering campaigns, political content, lobbying, and the ability to influence elections and behavior. 

Adjacently, Anthropic and OpenAI both have extensive Terms regarding the prohibited use of their services for criminal justice and surveillance applications. This includes specific Terms relating to the use of services to make decisions on criminal justice applications, tracking people (including emotional state and location---Google also has one clause relating to tracking/monitoring people without consent), and, interestingly, building systems that will infer the emotions of people. 

There is less emphasis on privacy violations; however, several companies have specific Terms that prohibit users from impersonating other humans (including children for grooming) or deceiving others with AI (Anthropic, OpenAI, and DeepSeek). In contrast, Google's Terms specify that users are not allowed to impersonate or create a fake persona, living or dead, to falsely attribute content or mislead others. In a similar vein, all had some language relating to fraud, deception, or scams, alongside the disruption of their or other services. This is sometimes related to phishing and spam, while other examples relate to more specific things, such as falsifying documents or financial abuse (e.g., debt collection abuse). 

Three companies have specific clauses stating that users cannot use their services for professional advice (e.g., medical, legal) without appropriate professional input (Anthropic, OpenAI, DeepSeek). Google's wider API policy also notes this, but it is not included in the prohibited AI usage policy. OpenAI also has a section in their overall `Service Terms' regarding medical use, stating: Our \textit{`Services are not intended for use in the diagnosis or treatment of any health condition. You are responsible for complying with applicable laws for any use of our Services in a medical or healthcare context.'} This becomes somewhat at odds with their new product ChatGPT Health, launched on January 7th, 2026 \cite{OpenAI2026-ui}, where they state: \textit{`Health is already one of the most common ways people use ChatGPT, with hundreds of millions of people asking health and wellness questions each week,' [...] `based on our de-identified analysis of conversations, over 230 million people globally ask health and wellness related questions on ChatGPT every week.'}

Similarly, Anthropic's Terms classify healthcare as a `High-Risk Use Case' requiring human-in-the-loop oversight and disclosure to users. However, in January 2026, Anthropic announced `Claude for Healthcare' with HIPAA-ready infrastructure and connectors to clinical databases including CMS, Medidata, and ClinicalTrials.gov \cite{Anthropic2026-er}, actively promoting healthcare applications including prior authorization reviews, patient care coordination, and clinical trial operations---representing a strategic shift toward commercializing a domain their Terms treat with heightened caution.

All but DeepSeek have specific requirements for `high-stakes' situations. For example, Anthropic: `High-Risk Use Case Requirements', where they state that users who are using their services for any high-risk applications are required to have humans-in-the-loop and also have adequate disclosure to those impacted. Examples of high-risk use cases include legal, healthcare, insurance, employment, housing, academic testing, media, and journalism applications. xAI has a single clause relating to `high-stakes' automation, noting items relating to safety, legal, material rights, and well-being: \textit{`financial credit, educational, employment, housing, insurance, legal, medical, or other important decisions about or for them'}. OpenAI also requires human review in high-stakes automation, including \textit{`critical infrastructure, education, housing, employment, financial activities and credit, insurance, legal, medical, essential government services, product safety components, national security, migration, and law enforcement.`} The examples of `high-stakes' scenarios were fairly unanimous across companies.

\subsection{Restrictive Terms for Research}
When reviewing the Terms of the different platforms, several insights arise. Perhaps the one causing the most difficulty for researchers (and users more generally) is how different these Terms are for each company. For example, some appear to be relatively specific, whereby they note the restriction of usage in particular domains (e.g., usage never for `intelligence' or `national security' applications), whereas others appear to be much more open and vague terminology. 

Specifically, OpenAI has some of the most stringent restrictions, stating that usage can never include an `\textit{assessment or prediction of the risk of an individual committing a criminal offense based solely on their personal traits or on profiling}' that would prevent work looking at online risk signals of offline violence and terrorism, for example, \citet{Brown2024-xt} if work wanted to use LLM modeling rather than their approach using LDA topic modeling and other machine learning methods. This links to an earlier Term from OpenAI that states `\textit{Everyone has a right to safety and security. So you cannot use our services for: [...] national security or intelligence purposes without our review and approval}'. This has implications for security-oriented researchers and research funding bodies. 

Other companies pose restrictions on building inferences about people more generally; for example, Anthropic states that one cannot \textit{`Build or support emotional recognition systems or techniques that are used to infer emotions of a natural person, except for medical or safety reasons'}. This has implications for research focusing on human inferences, specifically emotion detection and sentiment \cite{Zhang2024-en}, which is commonly used to understand human perception of the world around them (e.g., political stance, happiness/loneliness, health interventions). There is increasing interest in the `social layer' of AI, which would be hindered if researchers wanted to utilize Anthropic's models. 

OpenAI has limitations regarding human inference; people are entitled to privacy. Therefore, we do not allow attempts to compromise the privacy of others, including aggregating, monitoring, profiling, or distributing individuals’ private or sensitive information without their authorization. And, \textit{`you may never use our services for: [...] evaluation or classification of individuals based on their social behavior, personal traits, or biometric data (including social scoring, profiling, or inferring sensitive attributes); (5) inference regarding an individual’s emotions in the workplace and educational settings, except when necessary for medical or safety reasons}'. This has substantial implications for large swathes of (computational) social science research, which, for example, is often interested in how we can link online and offline behaviors and attributes (e.g., `social behavior' or `personal traits', e.g., personality) \cite{Lazer2020-il, Stachl2020-pg}. A lot of current work using social media data could be analyzed using LLMs to predict inferences about individuals and groups, which also would likely be at-risk--from predicting sentiment and emotion, through to understanding how online behavior may influence offline behavior and outcomes.

Anthropic, OpenAI, and DeepSeek have Terms related to impersonating humans (including convincing others that the chatbot is human). This has implications for psychological studies that examine the comparison between LLM responses and humans. Anthropic also has specific terms noting deceptive techniques, which creates some uncertainty with academic usage, given that it is not uncommon for psychology studies to use deception. 

Anthropic also states that one is not allowed to: \textit{`Develop a product or support an existing service that deploys subliminal, manipulative, or deceptive techniques to distort behavior by impairing decision-making'}. This clause may have implications for some social and psychological research that utilizes deception and persuasion approaches in studies using LLMs/chatbots using their services. 

\section{Discussion}
Here, we sought to examine the current Terms of Use/Service relating to various LLM models and their companies and provide a resource of cross-referenced and categorized Terms. This aims to highlight (1) key restrictions on usage for general users and (2) the implications of these restrictions for research. This is important given the increasing integration of AI systems, including LLMs, into daily life and research pipelines. It is critical to be aware of the Terms and whether any work is being done using these models that is close to or potentially in breach of these Terms. This has downstream implications for what happens if a breach occurs and how it is resolved (and whether this is even possible). It is important to note that the ability to find the appropriate Terms for companies is not always straightforward. For example, there are Gemini API terms specifically, alongside wider Google API terms, alongside a specific Generative AI Prohibited Use Policy, which makes it difficult for all users to navigate and know which Terms actually apply. Despite having a range of terms (including those beyond the ones listed here), the Terms from Google/Gemini tend to be broad, vague, and light touch.

It is clear that there is a shift towards the removal of accountability of companies generating new technology onto the users via the Terms \cite{Muldowney2025-mc}. This behavior, while not new, is important to notice, particularly in light of suicides being linked to OpenAI's models being used in lieu of therapy and psychological help \cite{Yousif2025-xu}. Early research has shown that LLM models vary substantially in their appropriateness in responding to mental health crises \cite{Arnaiz-Rodriguez2025-nt, Moore2025-eg}, which is worrying, especially given the huge number of daily users of ChatGPT for health-oriented assistance \cite{OpenAI2026-ui}. This highlights key questions around safeguarding, including the design of the models to `behave’ and respond in human-like language, the harms this may bring, and the lack of safeguards in place to stop models from giving medical (or other) advice (and/or being jailbroken to do this). This is alongside the reality that people may have no other options, hence turning to services such as LLMs to get any form of help. Although OpenAI has attempted to tackle these issues, the level of transparency remains low. For example, OpenAI states that they have a ‘Global Physician Network’ without additional information about these 300 physicians, simply noting \textit{`These experts found that the new model was substantially improved compared to GPT‑4o, with a 39-52'\% decrease in undesired responses across all categories.'}. Further information on how this was tested and evaluated and whether any other external validation took place is not openly available for scrutiny \cite{OpenAI2025-ny}.

\subsection{Implications for Research}
First, this analysis has shown that it is critically important for researchers (and institutions or companies) to be aware of these Terms and to understand the implications for their research. This causes additional workload for researchers or specific staff in the institution to monitor this to ensure that projects are sitting within the Terms, which would be a vast amount of responsibility and work. For example, security researchers using LLMs in their work may need to reconsider the use of any OpenAI models because of questions related to `intelligence' and `national security' applications without their review (noting that there was no direct way linked to engage in this process). This causes issues for (governmental) funders that may be funded via the military or other security services, which will then likely mean that any projects funded by them would not be allowed to use particular services. This also requires those who are making decisions on which projects to fund to be aware and up-to-date with various Terms for any and all platforms, which is unsustainable. 

Furthermore, the adaptability of research can potentially change direction quickly, should these Terms change. This type of risk management would be useful but is a large undertaking, especially when research funding is involved, where projects may be tied to delivering something specific. While retrospective policies would generally not be applied, it is important to be aware of changes when they happen and to be agile enough to adapt and ensure that, if this is funded work, the funders are able to absorb the changes that are required. 

In other areas, it is clear that research building inferences relating to emotion, which is a large research space, may be limited in the models they can use, which again highlights the requirement for researchers to regularly read and understand Terms. Similarly, when considering Terms relating to `profiling' and human assessments, questions arise regarding whether there is an exception for behavioral and psychological science research, where there is often a large interest in individual differences and the ability to predict them based on various data. This is difficult given the substantial interest in LLMs across the social sciences as part of the modeling pipelines. Having exceptions for being able to build fake personae for experiments would be useful, but researchers would be limited in this case from using certain models. Furthermore, there have been a number of papers published demonstrating how users can poison and how simple it is to breach guardrails, for example, \citet{Bowen2025-bd}, which is critically important research to understand the vulnerabilities in an independent research setting rather than only via the companies themselves. However, this kind of work, red-teaming or not, would be prohibited despite its importance, especially due to the widespread adoption and use of LLMs. 

Another issue may be for various AI and technology companies seeking to build AI or LLM systems. These Terms may be difficult to fit within if companies want to use out-of-the-box tools, such as OpenAI or Anthropic's models, as a benchmarking exercise.

\subsection{Implications for Policy}
\subsubsection{AI and LLM Harms}
While Terms are there to provide users awareness as to what they can and cannot do with a product or service, there remain questions as to how Terms are actually enforced. The main example at present is users of xAI's Grok using the service to generate non-consensual sexual imagery of children and women, and undressing women using their images. Creating and sharing explicit images of women has been stated as illegal by the European Commission \cite{Sandle2026-xj} and in the UK under the Online Safety Act \cite{Ministry-of-Justice2026-yb}. Specifically, using AI for the sexual exploitation of children and child pornography is illegal under the EU AI Act (Annex II) and UK law. Based on our findings, the images of children also breach the Terms given xAI's child safety Terms; however, there are no other Terms specifically relating to the use of their service for sexual violence or non-consensual intimate content of adults. This is also the case for Anthropic and DeepSeek. However, Anthropic has specific terms against the use of their services for graphic violence (including sexual) and `developing products or tools that facilitate deception and intent to cause emotional harm'---but it remains unknown how much protection these Terms and various regulations actually provide. However, as this has continued without real interventions, X could be banned in the UK, where Ofcom is launching investigations \cite{Field2026-hx}. While LLMs are still relatively young technologies, it is not known how far behind law and regulations are, but the route forward continues to be uncertain. 

Nothing has yet been done to stop users from engaging in this incredibly harmful and misogynistic behavior with Grok, and this may be replicated should this happen on other services, given the substantial variance in Terms and how breaches like this are caught and enforced. Here, it appears that users reported this issue as opposed to internal mechanisms and processes identifying this. This raises other issues regarding the recovery of the harms that have taken place to the victims, removal of the images, and the consequences for those who generated the images. This is particularly critical considering that Grok is used to generate child sexual abuse material (CSAM). 

Another issue that arises is that it continues to be simple for users to override the guardrails provided by companies. This is problematic in the case of Grok's image generation, but equally so for other cases, such as users engaging with LLMs to access emotional, therapeutic, and medical support. While some LLMs have disclaimers \cite{Wachter2024-ck}, it is not yet known whether this deters users. This generates a number of questions: First, how should LLMs respond when users approach guardrails, and what should the process be to prevent harm to the user or others? Second, how do we design user interfaces (UIs) to prevent users from engaging in harmful activities? Third, should there be convergence in Terms across companies to avoid harm, and how would this be conceptualized and executed, given the number of barriers, from language, culture, regulatory landscapes, to country cooperation? 

Furthermore, it is important that regulation development also considers their definitions and boundaries regarding what technologies are included under their frameworks. For example, platform data access is mandated in the EU DSA and UK Online Harms Act, but it remains unclear whether LLMs fall within these regulations and thus arguably fall outside these mandates. 

Alongside recent examples of harms to children and women, it is important to be aware of other harms directly relating to LLMs more generally, as noted by \citet{Weidinger2022-ql}, which include: \textit{`I. Discrimination, Exclusion and Toxicity, II. Information Hazards, III. Misinformation Harms, V. Malicious Uses, V. Human-Computer Interaction Harms, VI. Automation, Access, and Environmental Harms.'} While some Terms attempt to include clauses that respond to these issues, it is a long way from good enough to enforce and actually shape change and AI safety. 

\subsubsection{The Future of AI and LLM Research}
Terms continue to evolve, but this has downstream impacts on all users. Terms typically become stricter over time, which reduces their applicability and has implications for researcher access. It is important that policy considers this: independent research is critical for assessing and evaluating the impact of these systems; therefore, it is also critical to ensure access to these services for research purposes. This would be more appropriate than potential blanket applications, whereby no one can use services for particular domains. For example, there are pilots and early shifts into using AI systems for job interviews and policing interviews, among other sensitive areas, which should be independently evaluated and examined. However, the academic ability to test this using some models is currently blocked. 

We propose that policymakers and research institutions advocate for a standardized `Research Use' addendum to the Terms. Such a framework would explicitly permit \textit{bona fide} research activities (including those involving sensitive topics) with appropriate ethical oversight. Such an addendum could be modeled on existing academic licensing agreements for proprietary software and datasets, providing clear legal pathways for researchers while maintaining the companies’ safeguards.

\section{Limitations}
This analysis has some limitations. First, it represents a snapshot of the Terms as of November 2025 (no changes upon submission in January 2026); given the dynamic nature of these documents, the findings may shift over time. Second, our focus on five prominent LLM providers, while representative of current market leaders, does not capture the entire ecosystem of available models. Third, our identified implications for research would benefit from validation through interviews with affected researchers, which is an avenue for future research.

\section{Conclusion}
Overall, it appears that users of AI systems, including LLMs, need to become legal experts. This study aimed to examine the current Terms for various AI companies that distribute LLMs, and we found that there are a number of issues with the Terms for daily and research usage. It is critically important to ensure that systems are in place to protect independent research and users of AI systems, rather than continuing to allow companies to quietly shift accountability onto users, away from themselves, despite their continued development of technologies without appropriate safeguarding and responsibility. It is unfortunate that current legal and regulatory space inadvertently allows companies to continue to reduce their liability and shift accountability elsewhere, rather than tackling issues and harms head-on and making their services, products, and technologies safer, more trustworthy, and more truthful. 

\section{Endmatter Sections}
\subsection{Generative AI Usage Statement}
Generative AI was not used for the research process, from idea development and analysis to writing. However, the lead author used Gemini 3 to help improve and fix the formatting in \LaTeX prior to submission.

\subsection{Authorship Contributions}
\textbf{Conceptualization:} BID;
\textbf{Methodology:} BID;
\textbf{Formal Analysis \& Resource Development:} BID;
\textbf{Writing (Original Draft):} BID;
\textbf{Writing (Review \& Editing):} BID, KM, FADB, ANJ

\subsection{Acknowledgments \& Competing Interests}
No funding is associated with this work.

\bibliographystyle{abbrv}
\bibliography{bib}

@MISC{Anthropic2026-er,
  title        = "Advancing Claude in healthcare and the life sciences",
  author       = "{Anthropic}",
  month        = jan,
  year         = 2026,
  howpublished = "\url{https://www.anthropic.com/news/healthcare-life-sciences}"
}

@ARTICLE{Burgess2026-ew,
  title    = "Grok is pushing {AI} 'undressing' mainstream",
  author   = "Burgess, Matt and Varner, Maddy",
  journal  = "Wired",
  month    =  "6~" # jan,
  year     =  2026
}

@INPROCEEDINGS{Zhang2024-en,
  title     = "Sentiment analysis in the era of large language models: A reality
               check",
  author    = "Zhang, Wenxuan and Deng, Yue and Liu, Bing and Pan, Sinno and
               Bing, Lidong",
  booktitle = "Findings of the Association for Computational Linguistics: NAACL
               2024",
  publisher = "Association for Computational Linguistics",
  address   = "Stroudsburg, PA, USA",
  pages     = "3881--3906",
  year      =  2024,
  doi       = "10.18653/v1/2024.findings-naacl.246"
}

@ARTICLE{Stachl2020-pg,
  title     = "Predicting personality from patterns of behavior collected with
               smartphones",
  author    = "Stachl, Clemens and Au, Quay and Schoedel, Ramona and Gosling,
               Samuel D and Harari, Gabriella M and Buschek, Daniel and Völkel,
               Sarah Theres and Schuwerk, Tobias and Oldemeier, Michelle and
               Ullmann, Theresa and Hussmann, Heinrich and Bischl, Bernd and
               Bühner, Markus",
  journal   = "Proceedings of the National Academy of Sciences of the United
               States of America",
  publisher = "Proceedings of the National Academy of Sciences",
  volume    =  117,
  number    =  30,
  pages     = "17680--17687",
  month     =  "28~" # jul,
  year      =  2020,
  doi       = "10.1073/pnas.1920484117"
}

@INPROCEEDINGS{Moore2025-eg,
  title     = "Expressing stigma and inappropriate responses prevents {LLMs}
               from safely replacing mental health providers",
  author    = "Moore, Jared and Grabb, Declan and Agnew, William and Klyman,
               Kevin and Chancellor, Stevie and Ong, Desmond C and Haber, Nick",
  booktitle = "Proceedings of the 2025 ACM Conference on Fairness,
               Accountability, and Transparency",
  publisher = "ACM",
  address   = "New York, NY, USA",
  pages     = "599--627",
  month     =  "23~" # jun,
  year      =  2025,
  doi       = "10.1145/3715275.3732039"
}

@ARTICLE{Arnaiz-Rodriguez2025-nt,
  title         = "Between help and harm: An evaluation of mental health crisis
                   handling by {LLMs}",
  author        = "Arnaiz-Rodriguez, Adrian and Baidal, Miguel and Derner, Erik
                   and Annable, Jenn Layton and Ball, Mark and Ince, Mark and
                   Vallejos, Elvira Perez and Oliver, Nuria",
  journal       = "arXiv [cs.CL]",
  month         =  "2~" # dec,
  year          =  2025,
  archivePrefix = "arXiv",
  doi           = "10.48550/arXiv.2509.24857"
}

@ARTICLE{Steinfeld2016-nf,
  title     = "“I agree to the terms and conditions”: (How) do users read
               privacy policies online? An eye-tracking experiment",
  author    = "Steinfeld, Nili",
  journal   = "Computers in Human Behavior",
  publisher = "Elsevier BV",
  volume    =  55,
  pages     = "992--1000",
  month     =  feb,
  year      =  2016,
  doi       = "10.1016/j.chb.2015.09.038"
}

@ARTICLE{Bowen2025-bd,
  title         = "Scaling trends for data poisoning in {LLMs}",
  author        = "Bowen, Dillon and Murphy, Brendan and Cai, Will and
                   Khachaturov, David and Gleave, Adam and Pelrine, Kellin",
  journal       = "arXiv [cs.CR]",
  month         =  "17~" # jul,
  year          =  2025,
  archivePrefix = "arXiv",
  doi           = "10.48550/arXiv.2408.02946"
}

@ARTICLE{Bowen2009-nw,
  title     = "Document analysis as a qualitative research method",
  author    = "Bowen, Glenn A",
  journal   = "Qualitative Research Journal",
  publisher = "Emerald Publishing",
  volume    =  9,
  number    =  2,
  pages     = "27--40",
  month     =  "3~" # aug,
  year      =  2009,
  doi       = "10.3316/QRJ0902027"
}

@ARTICLE{Freelon2024-yi,
  title     = "The post-{API} age of social media data access: Past, present,
               and future",
  author    = "Freelon, Deen and Monzer, Cristina and Jeon, Gayoung and Moy,
               Cameron and Williams, Natasha",
  journal   = "The Annals of the American Academy of Political and Social
               Science",
  publisher = "SAGE Publications",
  volume    =  715,
  number    =  1,
  pages     = "16--37",
  month     =  sep,
  year      =  2024,
  doi       = "10.1177/00027162251372557"
}

@ARTICLE{Field2026-hx,
  title     = "Musk’s {X} could be banned in Britain over {AI} chatbot row",
  author    = "Field, Matthew and Holl-Allen, Genevieve and Stringer, Connor",
  journal   = "The Sunday Telegraph",
  publisher = "The Telegraph",
  month     =  "8~" # jan,
  year      =  2026
}

@MISC{Ministry-of-Justice2026-yb,
  title        = "Better protection for victims thanks to new law on sexually
                  explicit deepfakes",
  author       = "{Ministry of Justice} and Sackman KC MP, Sarah",
  booktitle    = "GOV.UK",
  month        =  jan,
  year         =  2026,
  howpublished = "\url{https://www.gov.uk/government/news/better-protection-for-victims-thanks-to-new-law-on-sexually-explicit-deepfakes}"
}

@MISC{OpenAI2026-ui,
  title        = "Introducing {ChatGPT} Health",
  author       = "{OpenAI}",
  booktitle    = "OpenAI - Product",
  month        =  jan,
  year         =  2026,
  howpublished = "\url{https://openai.com/index/introducing-chatgpt-health/}"
}

@ARTICLE{Gentleman2026-pt,
  title     = "Grok {AI} still being used to digitally undress women and
               children despite suspension pledge",
  author    = "Gentleman, Amelia and Horton, Helena and Milmo, Dan",
  journal   = "The Guardian",
  publisher = "The Guardian",
  month     =  "5~" # jan,
  year      =  2026
}

@ARTICLE{Sandle2026-xj,
  title     = "European Commission calls Grok's sexualised {AI} photos
               'illegal,' Britain demands answers",
  author    = "Sandle, Paul and Rasmussen, Louise",
  journal   = "Reuters",
  publisher = "Reuters",
  month     =  "5~" # jan,
  year      =  2026
}

@ARTICLE{Dhanji2025-vk,
  title     = "Deloitte to pay money back to Albanese government after using
               {AI} in \$440,000 report",
  author    = "Dhanji, Krishani",
  journal   = "The Guardian",
  publisher = "The Guardian",
  month     =  "6~" # oct,
  year      =  2025
}

@MISC{Paoli2025-oo,
  title        = "Deloitte allegedly cited {AI}-generated research in a
                  million-dollar report for a Canadian provincial government",
  author       = "Paoli, Nino",
  booktitle    = "Fortune",
  month        =  "25~" # nov,
  year         =  2025,
  howpublished = "\url{https://fortune.com/2025/11/25/deloitte-caught-fabricated-ai-generated-research-million-dollar-report-canada-government/}"
}

@INPROCEEDINGS{Kitkowska2022-gz,
  title     = "Online terms and conditions: Improving user engagement,
               awareness, and satisfaction through {UI} design",
  author    = "Kitkowska, Agnieszka and Högberg, Johan and Wästlund, Erik",
  booktitle = "CHI Conference on Human Factors in Computing Systems",
  publisher = "ACM",
  address   = "New York, NY, USA",
  month     =  "29~" # apr,
  year      =  2022,
  doi       = "10.1145/3491102.3517720"
}

@MISC{OpenAI2025-ny,
  title        = "Strengthening {ChatGPT’s} responses in sensitive conversations",
  author       = "{OpenAI}",
  booktitle    = "OpenAI",
  month        =  oct,
  year         =  2025,
  howpublished = "\url{https://openai.com/index/strengthening-chatgpt-responses-in-sensitive-conversations/}"
}

@MISC{Yousif2025-xu,
  title        = "Parents of teenager who took his own life sue {OpenAI}",
  author       = "Yousif, Nadine",
  booktitle    = "BBC",
  month        =  aug,
  year         =  2025,
  howpublished = "\url{https://www.bbc.co.uk/news/articles/cgerwp7rdlvo}"
}

@MISC{Muldowney2025-mc,
  title        = "{OpenAI} tries to shift responsibility to users",
  author       = "Muldowney, Decca and Bender, Emily M",
  booktitle    = "Mystery AI Hype Theater 3000: The Newsletter",
  month        =  nov,
  year         =  2025,
  howpublished = "\url{https://buttondown.com/maiht3k/archive/openai-tries-to-shift-responsibility-to-users/}"
}

@ARTICLE{Birhane2024-po,
  title     = "Large models of what? Mistaking engineering achievements for
               human linguistic agency",
  author    = "Birhane, Abeba and McGann, Marek",
  journal   = "Language Sciences (Oxford, England)",
  publisher = "Elsevier BV",
  volume    =  106,
  number    =  101672,
  pages     =  101672,
  month     =  nov,
  year      =  2024,
  doi       = "10.1016/j.langsci.2024.101672"
}

@ARTICLE{Xiong2024-sr,
  title         = "Everything everywhere all at once: {LLMs} can in-context
                   learn multiple tasks in superposition",
  author        = "Xiong, Zheyang and Cai, Ziyang and Cooper, John and Ge,
                   Albert and Papageorgiou, Vasilis and Sifakis, Zack and
                   Giannou, Angeliki and Lin, Ziqian and Yang, Liu and Agarwal,
                   Saurabh and Chrysos, Grigorios G and Oymak, Samet and Lee,
                   Kangwook and Papailiopoulos, Dimitris",
  journal       = "arXiv [cs.LG]",
  month         =  "7~" # oct,
  year          =  2024,
  archivePrefix = "arXiv"
}

@ARTICLE{Akhtarshenas2025-cu,
  title         = "{ChatGPT} or a silent everywhere helper: A survey of large
                   language models",
  author        = "Akhtarshenas, Azim and Dini, Afshin and Ayoobi, Navid",
  journal       = "arXiv [cs.CL]",
  month         =  "19~" # mar,
  year          =  2025,
  archivePrefix = "arXiv"
}

@ARTICLE{Zadushlivy2025-sh,
  title     = "Exploration of reproductive health apps' data privacy policies
               and the risks posed to users: Qualitative content analysis",
  author    = "Zadushlivy, Nina and Biviji, Rizwana and Williams, Karmen S",
  journal   = "Journal of Medical Internet Research",
  publisher = "JMIR Publications Inc.",
  volume    =  27,
  pages     = "e51517",
  month     =  "5~" # mar,
  year      =  2025,
  doi       = "10.2196/51517"
}

@INPROCEEDINGS{Dong2022-tn,
  title     = "Privacy analysis of period tracking mobile apps in the post-Roe
               v. Wade era",
  author    = "Dong, Zikan and Wang, Liu and Xie, Hao and Xu, Guoai and Wang,
               Haoyu",
  booktitle = "Proceedings of the 37th IEEE/ACM International Conference on
               Automated Software Engineering",
  publisher = "ACM",
  address   = "New York, NY, USA",
  pages     = "1--6",
  month     =  "10~" # oct,
  year      =  2022,
  doi       = "10.1145/3551349.3561343"
}

@INPROCEEDINGS{Luger2013-ru,
  title     = "Consent for all: revealing the hidden complexity of terms and
               conditions",
  author    = "Luger, Ewa and Moran, Stuart and Rodden, Tom",
  booktitle = "Proceedings of the SIGCHI Conference on Human Factors in
               Computing Systems",
  publisher = "ACM",
  address   = "New York, NY, USA",
  pages     = "2687--2696",
  month     =  "27~" # apr,
  year      =  2013,
  doi       = "10.1145/2470654.2481371"
}

@INPROCEEDINGS{Cao2024-mj,
  title     = "``{I} deleted it after the overturn of Roe v. Wade'':
               Understanding women's privacy concerns toward period-tracking
               apps in the post Roe v. Wade era",
  author    = "Cao, Jiaxun and Laabadli, Hiba and Mathis, Chase H and Stern,
               Rebecca D and Emami-Naeini, Pardis",
  booktitle = "Proceedings of the CHI Conference on Human Factors in Computing
               Systems",
  publisher = "ACM",
  address   = "New York, NY, USA",
  volume    =  18,
  pages     = "1--22",
  month     =  "11~" # may,
  year      =  2024,
  doi       = "10.1145/3613904.3642042"
}

@INPROCEEDINGS{Weidinger2022-ql,
  title     = "Taxonomy of risks posed by language models",
  author    = "Weidinger, Laura and Uesato, Jonathan and Rauh, Maribeth and
               Griffin, Conor and Huang, Po-Sen and Mellor, John and Glaese,
               Amelia and Cheng, Myra and Balle, Borja and Kasirzadeh, Atoosa
               and Biles, Courtney and Brown, Sasha and Kenton, Zac and Hawkins,
               Will and Stepleton, Tom and Birhane, Abeba and Hendricks, Lisa
               Anne and Rimell, Laura and Isaac, William and Haas, Julia and
               Legassick, Sean and Irving, Geoffrey and Gabriel, Iason",
  booktitle = "2022 ACM Conference on Fairness, Accountability, and Transparency",
  publisher = "ACM",
  address   = "New York, NY, USA",
  month     =  "21~" # jun,
  year      =  2022,
  doi       = "10.1145/3531146.3533088"
}

@INPROCEEDINGS{Amos2021-fp,
  title     = "Privacy policies over time: Curation and analysis of a
               million-document dataset",
  author    = "Amos, Ryan and Acar, Gunes and Lucherini, Eli and Kshirsagar,
               Mihir and Narayanan, Arvind and Mayer, Jonathan",
  booktitle = "Proceedings of the Web Conference 2021",
  publisher = "ACM",
  address   = "New York, NY, USA",
  month     =  "19~" # apr,
  year      =  2021,
  doi       = "10.1145/3442381.3450048"
}

@ARTICLE{Burnat2025-kr,
  title         = "The accountability paradox: How platform {API} restrictions
                   undermine {AI} transparency mandates",
  author        = "Burnat, Florian A D and Davidson, Brittany I",
  journal       = "arXiv [cs.CY]",
  month         =  "16~" # may,
  year          =  2025,
  archivePrefix = "arXiv"
}

@INPROCEEDINGS{Bender2021-ct,
  title     = "On the dangers of stochastic parrots: Can language models be too
               big?",
  author    = "Bender, Emily M and Gebru, Timnit and McMillan-Major, Angelina
               and Shmitchell, Shmargaret",
  booktitle = "Proceedings of the 2021 ACM Conference on Fairness,
               Accountability, and Transparency",
  publisher = "ACM",
  address   = "New York, NY, USA",
  pages     = "610--623",
  month     =  "3~" # mar,
  year      =  2021,
  doi       = "10.1145/3442188.3445922"
}

@ARTICLE{Brown2024-xt,
  title     = "Online signals of extremist mobilization",
  author    = "Brown, Olivia and Smith, Laura G E and Davidson, Brittany I and
               Racek, Daniel and Joinson, Adam",
  journal   = "Personality \& social psychology bulletin",
  publisher = "SAGE Publications",
  pages     =  1461672241266866,
  month     =  "31~" # jul,
  year      =  2024,
  doi       = "10.1177/01461672241266866"
}

@ARTICLE{Lazer2020-il,
  title     = "Computational social science: Obstacles and opportunities",
  author    = "Lazer, David M J and Pentland, Alex and Watts, Duncan J and Aral,
               Sinan and Athey, Susan and Contractor, Noshir and Freelon, Deen
               and Gonzalez-Bailon, Sandra and King, Gary and Margetts, Helen
               and Nelson, Alondra and Salganik, Matthew J and Strohmaier,
               Markus and Vespignani, Alessandro and Wagner, Claudia",
  journal   = "Science (New York, N.Y.)",
  publisher = "American Association for the Advancement of Science (AAAS)",
  volume    =  369,
  number    =  6507,
  pages     = "1060--1062",
  month     =  "28~" # aug,
  year      =  2020,
  doi       = "10.1126/science.aaz8170"
}

@ARTICLE{Davidson2023-uu,
  title     = "Platform-controlled social media {APIs} threaten open science",
  author    = "Davidson, Brittany I and Wischerath, Darja and Racek, Daniel and
               Parry, Douglas A and Godwin, Emily and Hinds, Joanne and van der
               Linden, Dirk and Roscoe, Jonathan F and Ayravainen, Laura and
               Cork, Alicia G",
  journal   = "Nature human behaviour",
  publisher = "Springer Science and Business Media LLC",
  volume    =  7,
  number    =  12,
  pages     = "2054--2057",
  month     =  "2~" # dec,
  year      =  2023,
  doi       = "10.1038/s41562-023-01750-2"
}

@ARTICLE{Wachter2024-ck,
  title     = "Do large language models have a legal duty to tell the truth?",
  author    = "Wachter, Sandra and Mittelstadt, Brent and Russell, Chris",
  journal   = "Royal Society open science",
  publisher = "The Royal Society",
  volume    =  11,
  number    =  8,
  month     =  aug,
  year      =  2024,
  doi       = "10.1098/rsos.240197"
}

\end{document}